\documentclass[lettersize,journal]{IEEEtran}
\usepackage{amsmath,amsfonts}
\usepackage{algorithmic}
\usepackage{algorithm}
\usepackage{array}
\usepackage[caption=false,font=normalsize,labelfont=sf,textfont=sf]{subfig}
\usepackage{textcomp}
\usepackage{stfloats}
\usepackage{url}
\usepackage{verbatim}
\usepackage{graphicx}
\usepackage{cite}
\usepackage{xcolor}
\begin{document}

\title{Radiation Total Dose for PRIMA: Cold Exposure with Alpha Particles}

\author{\IEEEauthorblockN{Elijah Kane\IEEEauthorrefmark{1}\IEEEauthorrefmark{2}, Chris Albert\IEEEauthorrefmark{1}\IEEEauthorrefmark{2}, 
Andrew Beyer\IEEEauthorrefmark{2}, Charles (Matt) Bradford\IEEEauthorrefmark{1}\IEEEauthorrefmark{2}, Pierre Echternach\IEEEauthorrefmark{2}, 
Logan Foote\IEEEauthorrefmark{1}\IEEEauthorrefmark{2}, Jason Glenn\IEEEauthorrefmark{3}, Henry (Rick) LeDuc\IEEEauthorrefmark{2}, 
Hien Nguyen\IEEEauthorrefmark{2}, Thomas Stevenson\IEEEauthorrefmark{3}, Brian Zhu\IEEEauthorrefmark{2},
and Jonas Zmuidzinas\IEEEauthorrefmark{1}\IEEEauthorrefmark{2}}
\\
\IEEEauthorrefmark{1}California Institute of Technology, 1200 E California Blvd, Pasadena, 91125, California, USA
\\
\IEEEauthorrefmark{2}Jet Propulsion Laboratory, California Institute of Technology, 4800 Oak Grove Dr, Pasadena, 91109, California, USA
\\
\IEEEauthorrefmark{3}NASA Goddard Space Flight Center, 8800 Greenbelt Rd, Greenbelt, 20771, Maryland, USA}
%\author{Elijah Kane,~\IEEEmembership{Staff,~IEEE,}
        % <-this % stops a space
%\thanks{This paper was produced by the IEEE Publication Technology Group. They are in Piscataway, NJ.}% <-this % stops a space
%\thanks{Manuscript received April 19, 2021; revised August 16, 2021.}}

\markboth{Journal of \LaTeX\ Class Files,~Vol.~14, No.~8, August~2021}%
{Shell \MakeLowercase{\textit{et. al.}}: A Sample Article Using IEEEtran.cls for IEEE Journals}

\IEEEpubid{0000--0000/00\$00.00~\copyright~2021 IEEE}

\maketitle

\begin{abstract}
The Probe far-Infrared Mission for Astrophysics (PRIMA) is a far-infrared (24 – 261 $\mu$m wavelengths)
 probe-class space observatory currently under Phase A study, which promises 
 orders-of-magnitude improvement in mapping speed over its predecessors. PRIMA will field exquisitely 
 sensitive (NEP $<$ 0.1 aW Hz$^{-1/2}$) kilopixel arrays of kinetic inductance detectors (KIDs) 
 for the Far-Infrared Enhanced Survey Spectrometer (FIRESS) instrument. PRIMA will orbit in 
 space at the Sun-Earth L2 point, where Planck found the
 energetic particle flux to be about 300/min/cm$^2$ \cite{planck2013}. 
 Thus, the possible effect of a high fluence of energetic 
 particles on the detector sensitivity must be characterized. Previous work has suggested that bombardment of KIDs 
 by ions can reduce the quasiparticle lifetime (Barends et. al. 2009 \cite{Barends2009}), but the 
 conditions of the experiment were not representative of a  
 detector which is continuously held at sub-Kelvin temperatures in the 
 energetic particle environment of L2 orbit.
 To better replicate the damage which would be produced by energetic particles  
 in this environment, we developed a fully cryogenic irradiation 
 experiment in which a stepper motor controls a screen which can block or reveal an alpha 
 particle emitter. This setup can be used to irradiate aluminum KID arrays fabricated for FIRESS 
 to well-controlled dose levels. In this work, we calculate the damage dose expected for a 5-year mission in 
 L2 orbit, and we irradiate an array to approximately 62$\%$ of this level.
 Before and after irradiation, we measure the quasiparticle lifetimes, resonant frequencies, 
 and quality factors of the detectors.
\end{abstract}
\begin{IEEEkeywords}
Far-infrared, kinetic inductance detectors, PRIMA, radiation hardness, alpha particles.
\end{IEEEkeywords}

\section{Introduction}

The far-infrared (far-IR) can roughly be defined as the wavelength range longwards of the James Webb 
Space Telescope and shortwards of the Atacama Large Millimeter Array ($\sim 25-300$ $\mu$m). 
Since the Herschel Space Observatory ceased observations in 2013, there have been no 
astronomical observatories operating within this wavelength range. 
A far-infrared space observatory with a cryogenically cooled primary mirror and sufficiently 
sensitive detectors would provide a 5 to 6 order-of-magnitude increase in mapping
speed over Herschel \cite{glenn_prima}. Such an observatory
would yield a wealth of information on the buildup of dust and metals in the universe over cosmic 
time, the coevolution of star formation and active galactic nucleus accretion over cosmic time, 
the origins of water ices in protoplanetary disks, and many more astrophysical questions that 
are best studied in the far-IR.

The Probe far-Infrared Mission for Astrophysics (PRIMA) \cite{glenn_prima} is a 
probe-class space observatory currently under Phase A study, which will perform hyperspectral imaging 
at 25 - 84 $\mu$m, polarimetry at 80 - 261 $\mu$m, 
and spectroscopy at 25 – 235 $\mu$m. PRIMA will have a cryogenically 
cooled primary mirror and exquisitely 
sensitive kilopixel arrays of kinetic inductance detectors (KIDs) 
to achieve performance that is background-limited by the photon shot noise from zodiacal and galactic dust.
The Far-Infrared Enhanced Survey Spectrometer (FIRESS) \cite{bradford_firess} instrument will perform moderate-resolution 
($\delta \lambda / \lambda \sim 100$) spectroscopy, which will require detector noise equivalent
powers (NEPs) at or below 0.1 aW Hz$^{-1/2}$.
To achieve these sensitivities, PRIMA will use lumped-element KIDs (LEKIDs) 
with an interdigitated capacitor (IDC) made of niobium and a 
thin-film inductor made of aluminum. The Al inductor is designed with a small 
volume ($V \sim 15$ $\mu$m$^3$) to maximize the detector's responsivity to 
absorbed power. The aluminum film is also kept as clean as possible during deposition to 
maximize the lifetime of quasiparticle excitations ($\tau_{qp}$) to increase the 
detectors' sensitivity. The KID's sensitivity is related to the  volume and the 
quasiparticle lifetime through the proportionality NEP $\propto V/\tau_{qp}$.

PRIMA will orbit the Sun-Earth second Lagrange point (L2). The 100 mK bolometers in the 
Planck High-Frequency Instrument (HFI) provided a measure of the impact of the L2 proton spectrum
on sensitive cryogenic devices, finding a flux of 
about 300 events/min/cm$^2$ \cite{planck2013}. Potential concerns for the PRIMA KIDs in this
environment include both long-term 
degradation due to total dose, and glitches in the signal stream.   
This work addresses the first concern (total dose); the signal-stream impacts
have been studied in Karatsu et. al. \cite{karatsu2019} and Kane et. al. \cite{kaneltd20}, and found to be manageable
with suitable mitigation measures.

\begin{figure}[!h]
\centering
\includegraphics[width=3in]{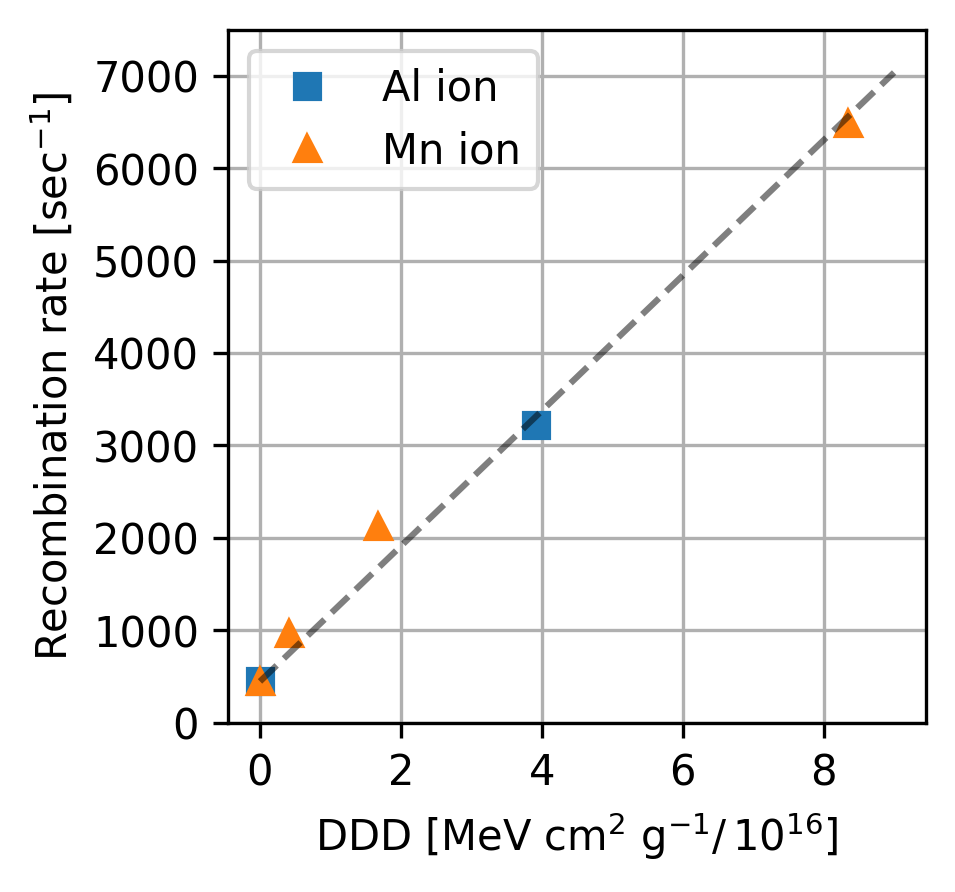}
\caption{Reduction in the quasiparticle lifetime of aluminum as a result of bombardment by Al and Mn ions. 
Plotted on the y-axis is the recombination rate, which is the inverse of quasiparticle lifetime. 
The data for ion energies, implanted ion concentrations, and quasiparticle lifetimes were taken from 
Barends et. al. 2009 \cite{Barends2009}. The methods described in Section \ref{sec:DDD background} 
were used to convert the ion energies and concentrations to displacement damage doses (DDD).}
\label{fig:Barends}
\end{figure}

In terms of total dose, the anticipated effect is degradation of the thin aluminum film 
due to the protons fluence. Previous work by Barends et. al. \cite{Barends2009} has shown
that bombardment of KIDs by Al and Mn ions can reduce the quasiparticle lifetime
(see Fig. \ref{fig:Barends}).
The proposed mechanism for lifetime reduction is the creation of disorder in the metallic
lattice within the aluminum inductor. Lattice 
defects may act as potential wells where quasiparticles can become 
trapped, leading to a residual population of quasiparticles which increase the recombination rate. 
The amount of lattice damage can be quantified using the displacement damage dose (DDD).

As we will discuss in Section \ref{sec:DDD background}, 
the DDD we expect to incur in L2 orbit is about 6 orders of magnitude less than the 
damage that was needed to reduce the quasiparticle lifetime in the Barends
experiment. However, we cannot 
assume that the results of the Barends experiment are a good proxy for irradiation in L2 orbit. 
First, the dominant source of DDD in L2 orbit will be from solar protons with energies of $10-100$ MeV, 
rather than the Al and Mn ions used in the Barends experiment. To produce a more accurate 
representation of the type of damage that PRIMA's KIDs will incur in L2 orbit, it would be best to 
irradiate the KIDs with protons or another particle with similar mass and charge.
Second, the ion bombardment in the Barends experiment was done at room temperature. 
It is possible that lattice defects in aluminum are mobile at room temperature, which would 
lead to possible annealing of radiation-induced damage during the irradiation, during the 
transport of the KID to the dilution refrigerator, and during the cooldown time. 
Thus, the actual amount of displacement damage necessary to decrease the 
quasiparticle lifetime may be lower than we would calculate from the data of 
Barends et. al.

In this work, we present a fully cryogenic irradiation experiment in which FIRESS KIDs can be 
exposed to controlled doses of alpha particles. As a complement to this work, we note an article 
in preparation by PRIMA team members analyzing multi-year equivalent exposure of a 100 mK KID array 
to a beam of $\sim 60$ MeV protons, followed by device characterization in the same cryogenic run (Karatsu et. al., in prep).

\section{Displacement damage dose in L2 orbit}
\label{sec:DDD background}

The energetic particle flux at L2 is composed of protons, electrons, and helium nuclei
from solar and galactic sources. 
In flight, the detectors will be shielded from these particles on all sides by an aluminum
metering structure and by the aluminum detector enclosure. These structures block all helium nuclei,
and we ignore electrons when calculating our mission DDD, as they contribute negligible DDD compared
to the protons. Thus, we only consider protons in our DDD calculation.
For the incident proton fluxes, we use the JPL Solar Proton model \cite{jpl_solar} for solar protons, and
the CREME96 model \cite{creme96} for galactic protons. To provide a conservative worst-case estimate
of the mission proton fluence, we evaluate the solar proton fluxes at the period of maximum
solar activity. We also evaluate the CREME96 model at a period of minimum solar activity,
when the flux of galactic protons at L2 is at a maximum.

The protons will lose energy as they pass through the aluminum surrounding the detectors.
The thickness of the metering structure varies with direction, from a few mm to $>200$ mm.
We model the metering structure as an aluminum box with walls having thicknesses listed in 
Table \ref{tab:thicknesses}.

\begin{table}[h]
\centering
\begin{tabular}{l l}
\hline
Direction & Thickness [mm] \\
\hline
+X & 11.80 \\
-X & 18.21 \\
+Y & 25.99 \\
-Y & 2.51 \\
+Z & 21.12 \\
-Z & 1.45 \\
\hline
\end{tabular}
\caption{The thicknesses of the walls used to model the aluminum metering structure.}
\label{tab:thicknesses}
\end{table}

\begin{figure}[!h]
\centering
\includegraphics[width=2.5in]{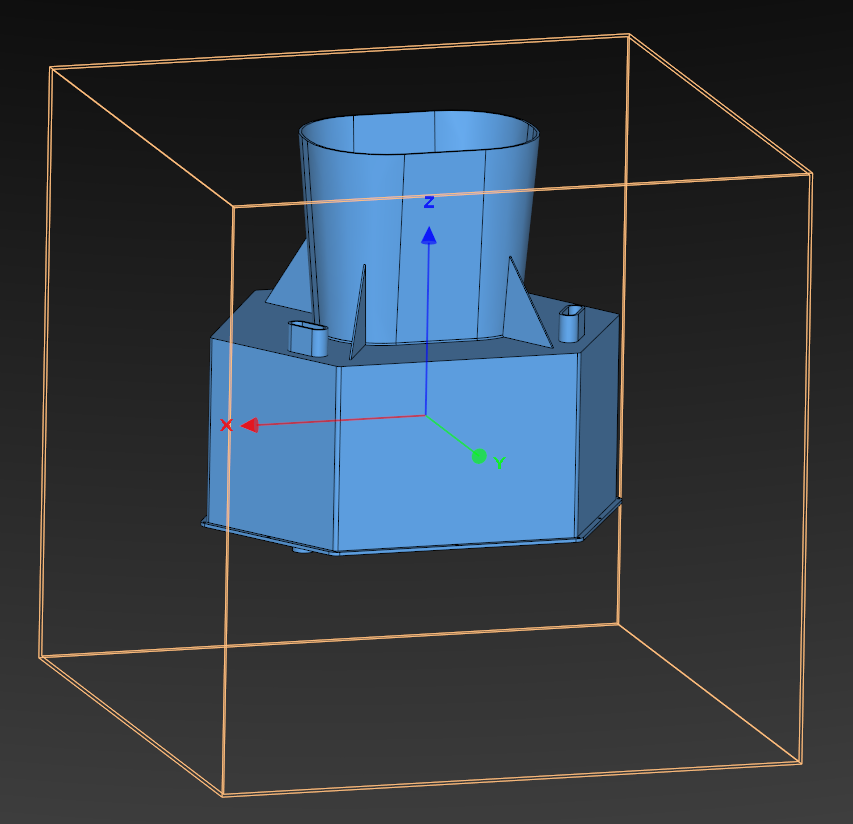}
\caption{Rendering of the aluminum detector enclosure. The orange box shows the location of the 
aluminum box used to model the metering structure surrounding the detector enclosure.}
\label{fig:ddd_sim}
\end{figure}

\begin{figure}[!h]
\centering
\includegraphics[width=3in]{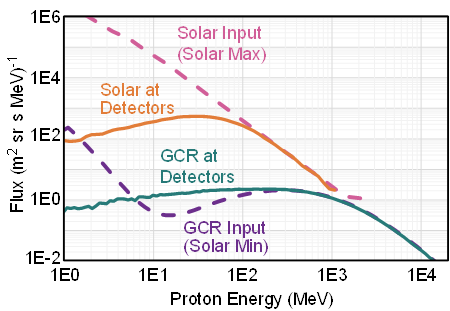}
\caption{Simulated solar and galactic cosmic ray (GCR) proton fluxes in Earth-Sun L2
orbit, calculated at solar maximum and solar minimum respectively to provide a 
conservative worst-case estimate of proton flux. Dashed lines correspond to no 
shielding, and solid lines correspond to our box-like shielding model.}
\label{fig:fluence}
\end{figure}

The detector enclosure and its location within the box are shown in Fig. \ref{fig:ddd_sim}.
The geometry of the detector enclosure is taken directly from a rendering of the spacecraft,
and it is not simplified in our model.
The input proton fluxes are transported through the box and the enclosure using GEANT4 \cite{geant4}.
Protons with initial energies less than $\sim 100$ MeV 
are found to be blocked by the shielding, protons with 
initial energies slightly above 100 MeV pass through with their energies significantly
reduced, and protons with energies well above 100 MeV pass through with the almost all of their initial energy.
The resulting proton flux spectra inside the detector enclosure 
is given by the solid lines in Fig. \ref{fig:fluence}.

The proton fluence $\Phi(E)$ at the detectors is the proton flux times the mission lifetime,
which we took to be 5.3 years. We combine the fluence with the non-ionizing energy loss
(NIEL) of protons in aluminum \cite{insoo_niel} to calculate the mission DDD, using Eq. (\ref{eq:ddd}).
The NIEL here is the energy lost by an incident proton which goes into displacing
aluminum atoms in the KID inductor from their normal lattice positions. Note that Eq. (\ref{eq:ddd}) assumes 
that the incident particle loses a negligible fraction of its energy as
it passes through the material, which is the case for both the L2 protons 
and the lab alpha particles considered in this work because the aluminum film making up the
inductor is very thin (30 - 40 nm). We obtain a mission DDD of
$2.9 \times 10^7$ MeV g$^{-1}$ for solar protons, and $6.0 \times 10^5$ MeV g$^{-1}$ for galactic protons,
for a total of $2.9 \times 10^7$ MeV g$^{-1}$.

\begin{equation}
\label{eq:ddd}
  \text{DDD} = \int \Phi(E) \text{NIEL}(E) dE.
\end{equation}

\section{Experimental methods}

To irradiate our KIDs with controllable dose levels, we designed a copper 
screen which can be moved by a stepper motor to block or reveal an alpha 
particle source near the KIDs. The stepper motor is from Osmtec \cite{osmtec}
(Part. No. 17HS15-0404S), and the ZK-SMC02 CNC stepper motor driver \cite{driver} 
was used to control the motor. This motor is suitable for use at temperatures near 
4 K if cryogenically-stable ball bearings are used. For example, the motor has been 
successfully used to operate a mechanical heat switch \cite{heatswitch}. 
We replaced the manufacturer's ball bearings, which included a nylon cage, with 
fully-stainless steel bearings. The bearings were taken apart, cleaned in 
an ultrasound bath with acetone for 1 hour, and then reassembled. We did not 
need to add a cryogenically-safe dry lubricant to make the 
motor work cryogenically.

The test was performed using a BlueFors LD400 dilution refrigerator 
\cite{bluefors} pre-cooled by a Cryomech PT415-RM pulse tube cryocooler 
\cite{cryomech}. The coldest stage, containing the detectors and the alpha
particle source, is cooled to a temperature of 10 mK. The cold stage was stable to
between 10 mK and 12 mK throughout the duration of the measurements, which will not 
affect the quasiparticle lifetimes, as the lifetimes are found to be constant
at temperatures below $\sim 150$ mK.
We place the stepper motor on the quasi-4 K stage rather than the cold stage for two reasons.
First, the motor's rotor is magnetic, and magnetic fields near the detectors
can shift their resonant frequencies \cite{vaughan2025}, introducing spurious signals. Thus,
the motor must be located far away from the detectors.
Second, the motor dissipates significant 
power when it is running. The current supplied to the motor when it is running
is 0.5 A, and the resistance of the copper wire carrying the 
current is about $5\,\Omega$. Thus, the power dissipated in the wire is on the order
of 1 W when the motor is running. This amount of power would significantly heat up 
the cold stage, which has a cooling power of $14\,\mu$W at a temperature of 20 mK
\cite{bluefors}. However, the Cryomech PT415-RM cryocooler, which cools the quasi-4 K stage,
provides a cooling power of 1.5 W at a temperature of 4.2 K \cite{cryomech}.
We found that the cryocooler could easily keep the quasi-4 K stage at around 3 K even 
when the motor was rotated by a half turn every 10 seconds, suggesting that 
1 W is a conservative estimate for the amount of power being dissipated. 
A nylon thread was tied to the 
rotor shaft, and feedthroughs blackened with Stycast 2850FT \cite{stycast} 
were used to run the thread down to the cold stage, where it was attached 
to the copper screen. 

\begin{figure}[!h]
\centering
\includegraphics[width=2.5in]{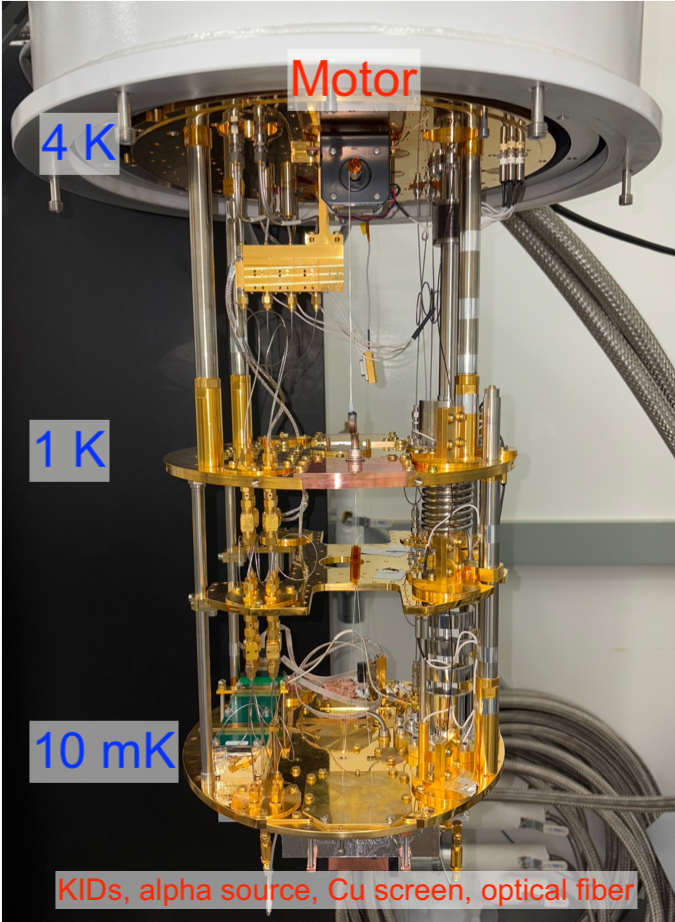}
\caption{The orientation of the stepper motor and the device under test within the 
dilution refrigerator. The motor is clamped to the 4 K stage, while the device, alpha 
source, copper screen, and optical fiber are positioned at the 10 mK stage.}
\label{fig:fridge}
\end{figure}

\begin{figure}[!h]
\centering
\includegraphics[width=3.5in]{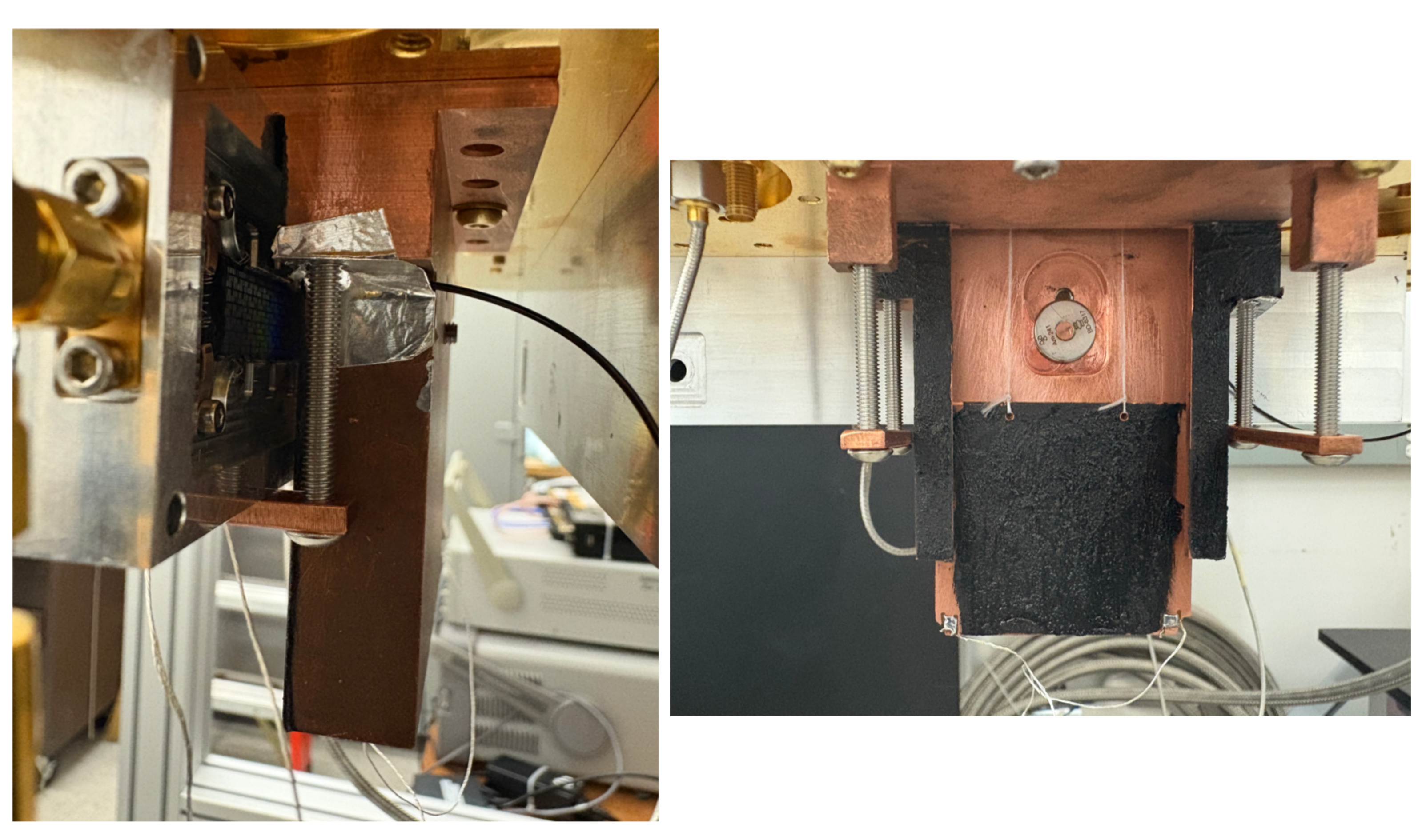}
\caption{\textit{Left:} The orientation of the optical fiber relative to 
the device under test. \textit{Right:} The view of the alpha particle source seen by the 
device when the copper screen is lowered.}
\label{fig:device}
\end{figure}

\begin{figure}[!h]
\centering
\includegraphics[width=3.5in]{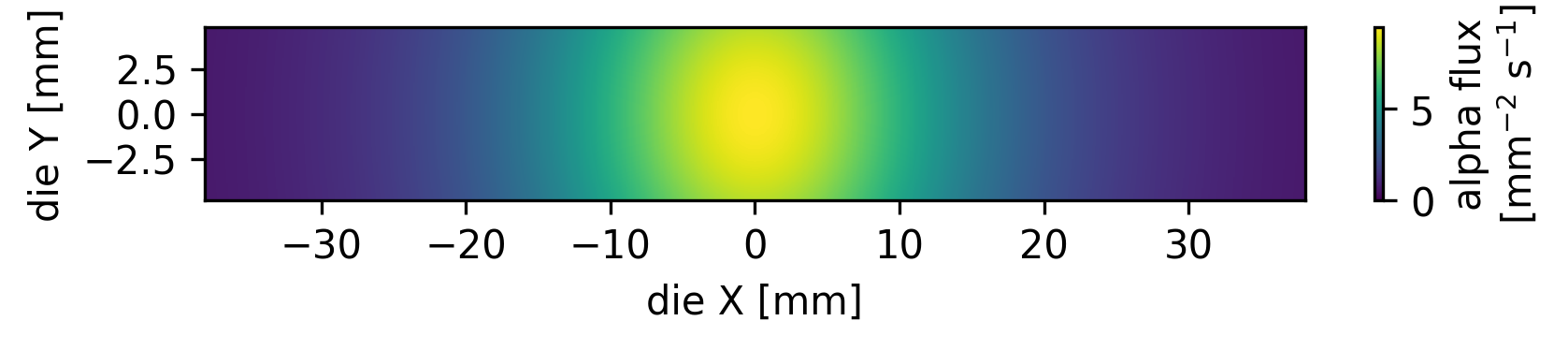}
\caption{A map of the modeled alpha particle fluxes incident on the device under test 
when the copper screen is lowered.}
\label{fig:flux map}
\end{figure}

\begin{figure}[!h]
\centering
\includegraphics[width=3in]{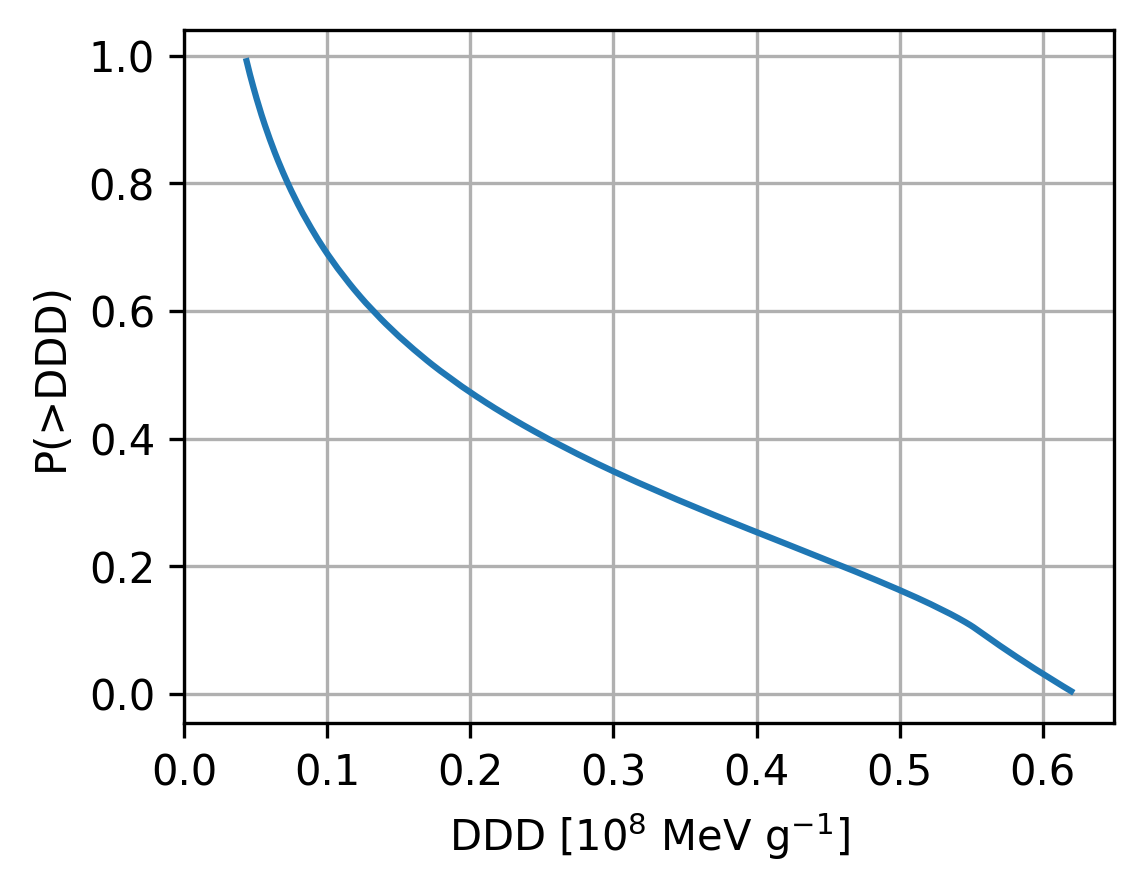}
\caption{The cumulative density function of displacement damage doses (DDD) experienced 
by KIDs across the device under test.}
\label{fig:DDD cdf}
\end{figure}

The alpha particle source was specially designed by Eckert and Ziegler \cite{enz} 
for use at sub-Kelvin temperatures. The active element is a film of Am-241, 
which emits alpha particles with energies of 5.486 MeV (85$\%$), 5.443 MeV 
(13$\%$), and 5.388 MeV (2$\%$). Our source has a measured activity of 
34.3 kBq. The source was placed 17 mm away from the center of our array. 
Treating it as a point source, this yields the alpha particle flux map shown 
in Fig. \ref{fig:flux map}, with a peak flux of $\sim 10$ mm$^{-2}$ s$^{-1}$.
We use the SR-NIEL-7 calculator \cite{srniel7} to obtain NIEL values for alpha particles
with $E=$ 5.486 MeV traversing through aluminum. As in Section \ref{sec:DDD background},
we use Eq. (\ref{eq:ddd}) to calculate the DDD.
In Fig. \ref{fig:DDD cdf}, we plot the 
cumulative density function of DDD values across the array for an irradiation time of 97 hours, 
which was the total irradiation time used in our experiment. The median DDD is 
$1.8\times 10^7$ MeV g$^{-1}$, which is 62$\%$ of the mission dose 
calculated in Section \ref{sec:DDD background}. The full range of DDDs is 
$(4.13\times 10^6, 6.20\times 10^7)$ MeV g$^{-1}$.

Before irradiating the device, we used a Control and Readout System (CRS) board from 
t0.technology \cite{Montgomery2024} to perform a forward transmission ($S_{21}$) 
sweep of the device and identify 
resonance features. We then performed narrower $S_{21}$ sweeps of all resonances
 across a range of readout power values, 
and fitted each KID's resonance feature at each power to a model incorporating 
the nonlinearity of the kinetic inductance with the current in the inductor \cite{swenson}. 
The model used for fitting is described in more detail in \cite{kaneltd20}. 
We selected the optimal power for each KID as the highest power at which the 
nonlinearity parameter $a$ was below $0.5$. This helps suppress the noise of 
the first-stage amplifier relative to the phase signal of the photon pulses, while 
still keeping the KID in the non-bifurcated regime.

Before and after the irradiation, an optical-fiber coupled laser oriented as in
Fig. \ref{fig:device} was used to illuminate the KIDs with $\lambda= 1550$ 
nm photons. We used a single-tone homodyne readout system to 
sequentially read out each KID for 60 seconds at a sample rate of 50 kHz. 
A simple peak-finding algorithm was used to identify photon pulses in each 
timestream. Pulses with overlapping tails (defined as pulses less than $3$ ms apart) 
and pulses with heights more than 3 standard deviations from the mean pulse height 
were rejected, and a template pulse was created for each KID by averaging the remaining pulses.
The pulses were then aligned using an optimal filter using the template pulse, 
and then a final mean pulse was created by averaging the pulses after 
alignment. A single-pole exponential 
was fitted to the mean pulse, starting at the point where the pulse had 
decayed to 0.6 times its maximum height, and the time constant of this 
fit was interpreted as the quasiparticle lifetime $\tau_{qp}$.

\section{Results}

From 557 resonance features which were identified from a frequency sweep 
of the device's forward transmission $S_{21}$, 312 KIDs were included in 
the results of this section. The other KIDs were left out due to poor 
calibration of the noise data from the IQ plane to a phase about the center 
of the IQ circle of the resonance. The most common cause of failed calibrations 
was the range of the amplifier 
noise being large compared to the radius of the IQ circle, making it not 
possible to accurately convert the noise IQ signal to a phase signal.

\begin{figure}[!h]
\centering
\includegraphics[width=3.5in]{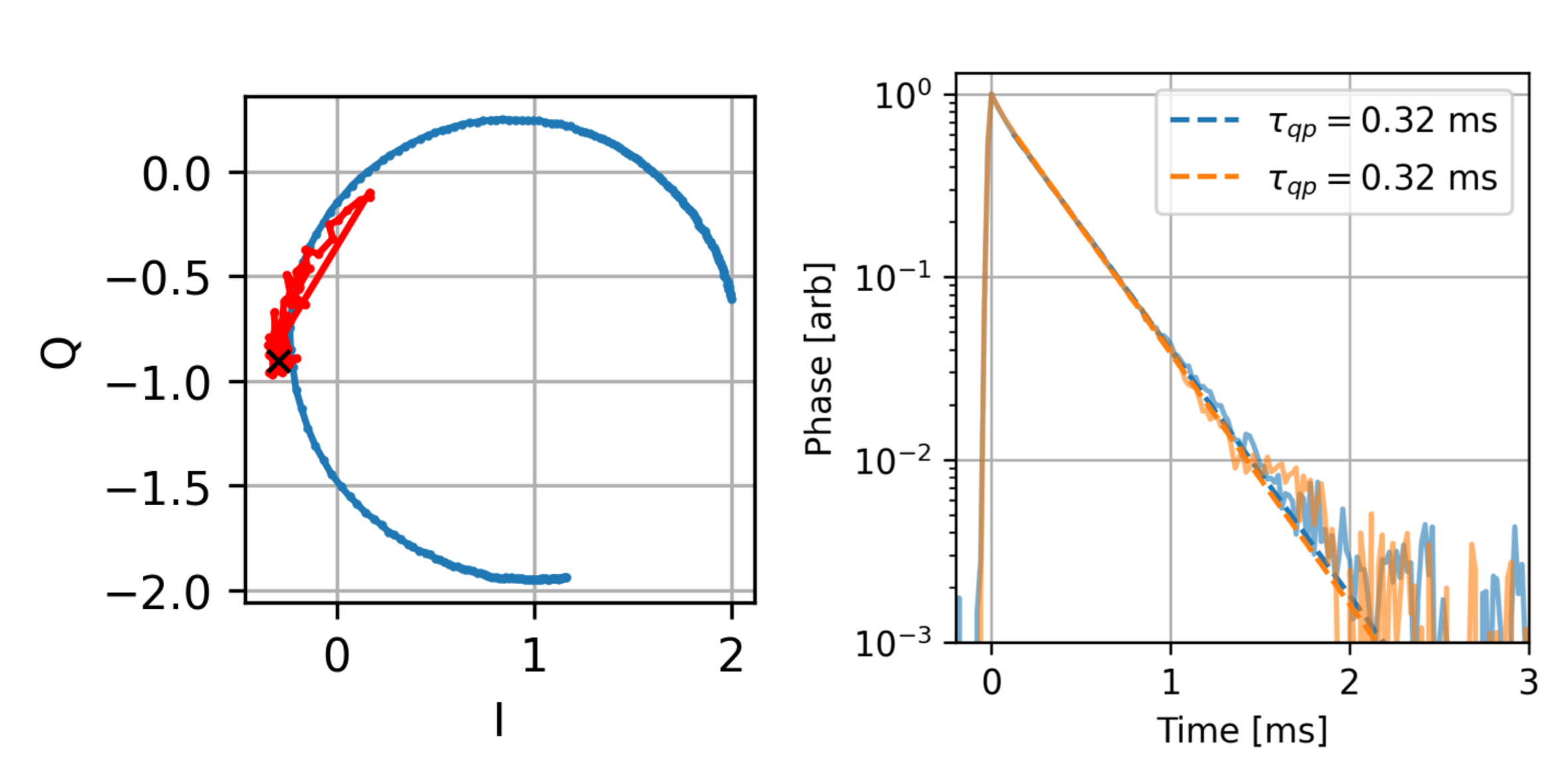}
\caption{\textit{Left:} A single photon pulse, plotted in the IQ plane over 
the resonance circle of the KID. The `x' shows the mean of the full IQ timestream, 
where we define phase $=0$. \textit{Right:} Single-pole exponential fits to 
the mean pulses for the same KID before and after irradiation, normalized to a 
peak amplitude of 1.}
\label{fig:pulse}
\end{figure}

Examples of mean pulses for the same KID before and after 
irradiation are shown in Fig. \ref{fig:pulse}, showing that the 
single-pole exponential function provides a good fit to the data. 
The pulses were kept in units of phase around the center of the 
resonance circle in the IQ plane for all signal processing steps.

\begin{figure}[!h]
\centering
\includegraphics[width=3in]{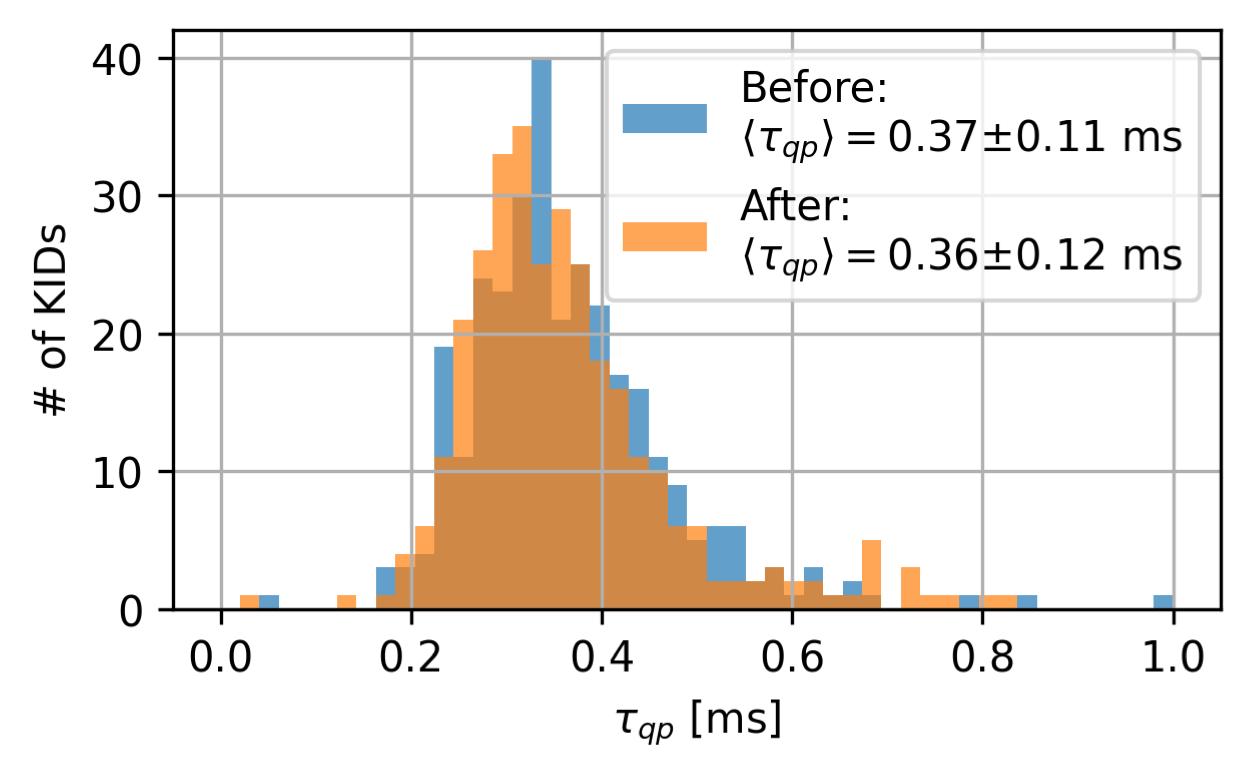}
\caption{Histogram of quasiparticle lifetimes before and after irradiation. 
The numbers in the legend are the mean and standard deviation of the distibutions.}
\label{fig:tau hist}
\end{figure}

\begin{figure}[!h]
\centering
\includegraphics[width=3in]{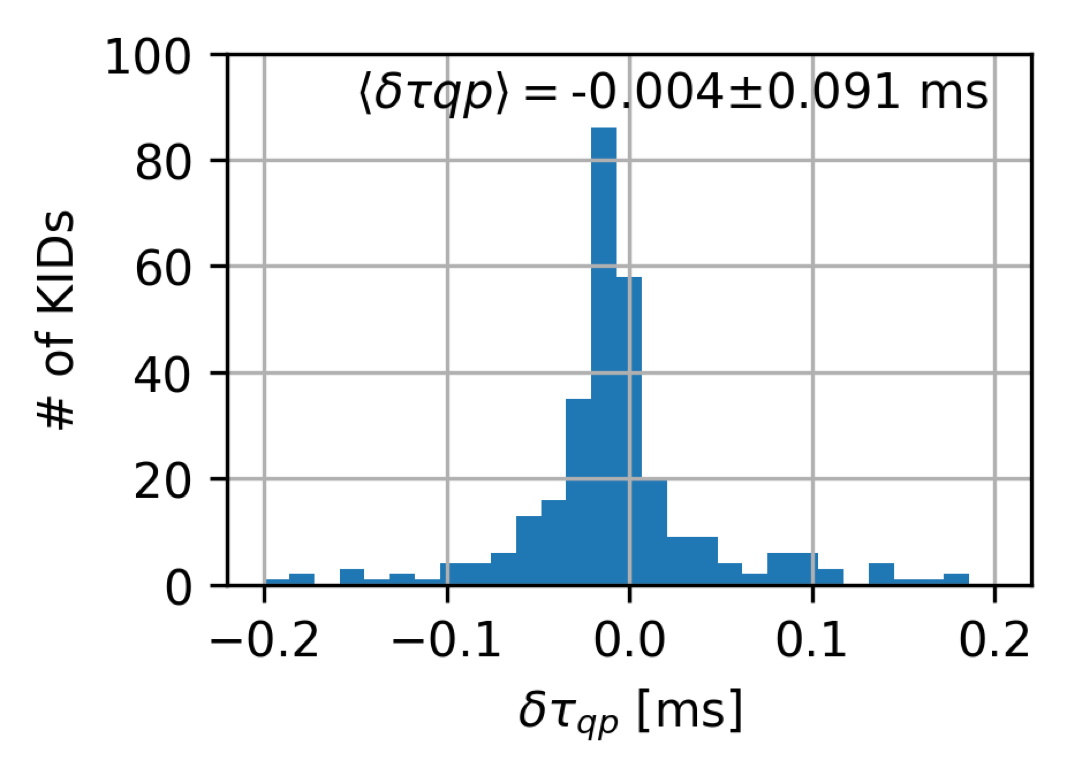}
\caption{Histogram of shifts in the quasiparticle lifetimes of each KID before 
and after irradiation.}
\label{fig:tau shift}
\end{figure}

In Fig. \ref{fig:tau hist}, we plot the histograms of all fitted lifetimes 
before and after the irradiation. There is a slight shift down in the mean 
value of $\tau_{qp}$ from 0.37 ms before irradiation to 0.36 ms after 
irradiation. In Fig. \ref{fig:tau shift}, we compare the shift in $\tau_{qp}$ 
for each pixel individually. There is a small mean downwards shift of 
$\langle \delta \tau_{qp} \rangle = -0.004$ ms.

\begin{figure}[!h]
\centering
\includegraphics[width=3in]{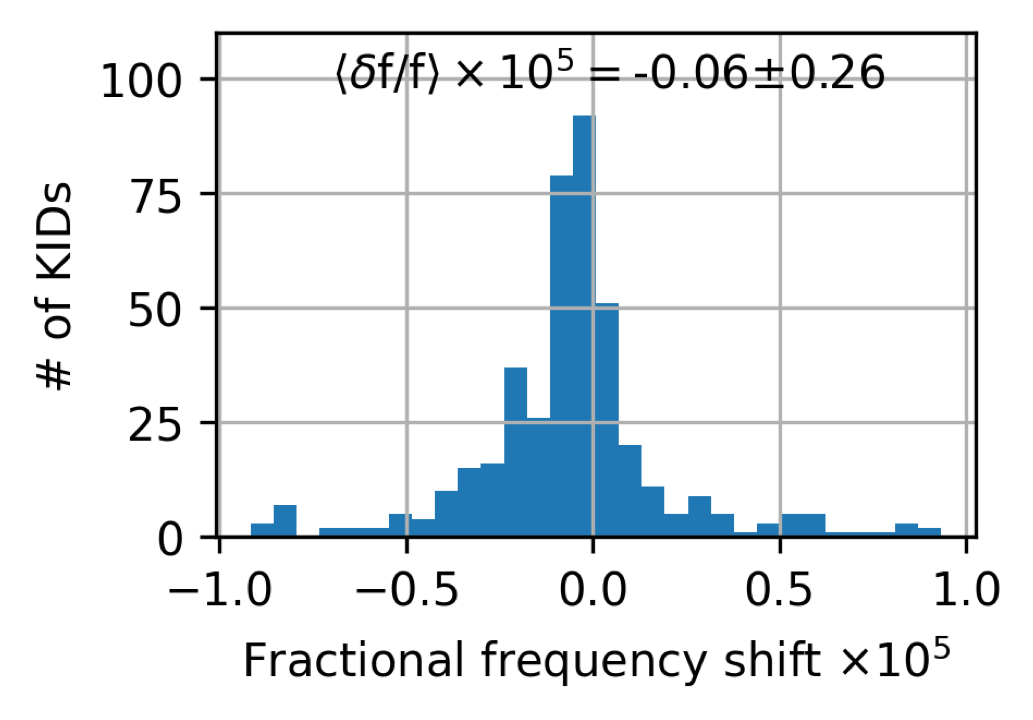}
\caption{Histogram of fractional frequency shifts of each KID before 
and after irradiation.}
\label{fig:freq shift}
\end{figure}

\begin{figure}[!h]
\centering
\includegraphics[width=3in]{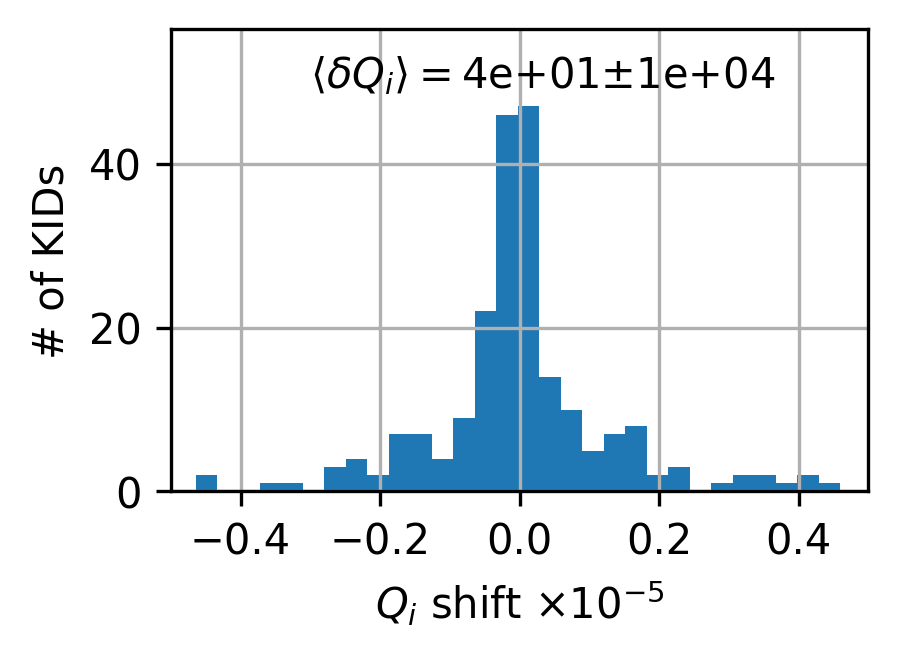}
\caption{Histogram of shifts in the internal quality factors of each KID before 
and after irradiation.}
\label{fig:Qi shift}
\end{figure}

We also measure the resonant frequency $f_r$ and internal quality factor 
$Q_i$ before and after irradiation through a fit to the forward transmission 
$S_{21}$ of each KID. 
The results for the fits are shown in Figs. \ref{fig:freq shift} and 
\ref{fig:Qi shift}. Small shifts of $-6$ kHz and $40$ are observed in the 
means of the distributions,
$\langle \delta f_r/f_r \rangle$ and $\langle Q_i \rangle$ respectively. 
In the histogram of $Q_i$ values, we excluded KIDs for which $Q_i$ 
was greater than $10$ times the coupling quality factor $Q_c$, because the 
fit value for $Q_i$ has a large scatter in the limit $Q_i/Q_c \gg 1$.

\section{Discussion}

The small negative values of $\langle \delta \tau_{qp} \rangle$ and 
$\langle \delta f_r \rangle$ are well within the measurement
errors given by the spread of the distributions of $\delta \tau_{qp}$ and 
$\delta f_r$. Although we do not believe that these shifts
are caused by the irradiation, we evaluate a worst-case scenario in which the shifts 
are attributed to degradation of the Al and/or Nb films which make up the KIDs.
In this case, we would expect the shift in $\tau_{qp}$ to scale up by $1/0.62 = 1.6 \times$ 
for the full L2 damage dose, from $\delta \tau_{qp} = -0.004$ ms to 
$\delta \tau_{qp} = -0.0065$ ms.
The most sensitive KID fabricated and measured for FIRESS had a limiting NEP of
$4.6\times 10^{-20}$ W Hz$^{-1/2}$ at 1 Hz, and a quasiparticle lifetime of $1.02$ ms 
\cite{Day2024}. Since the NEP is related to the quasiparticle lifetime by 
NEP $\propto\,\tau_{qp}^{-1}$, the NEP after such a decrease in the quasiparticle 
lifetime would be $4.63\times 10^{-20}$ W Hz$^{-1/2}$, which is still well below the 
goal of NEP $<$ 0.1 aW Hz$^{-1/2}$.

Assuming that the shifts and the scatter between the resonant frequencies 
before and after irradiation was caused by damage to the films, we can assume the 
scatter scales linearly with the DDD to estimate how much scatter would be present 
at the full L2 dose. The scatter is more important than the mean shift, because 
if resonators shift differently, they may collide with each other, decreasing 
the usable number of pixels in the array. Multiplying the 
observed scatter by $1.6\times$ gives a value of $4.2 \times 10^{-6}$. The arrays for 
FIRESS are designed to have resonant frequencies which are roughly equally separated 
in logarithmic space, with the lowest frequency being $500$ MHz, the highest frequency 
being $2,500$ MHz, and the number of pixels being $1,008$. Each resonator will 
be higher or lower than its neighbor by $\delta f_r/f_r \sim 1.6\times 10^{-3}$, which
is $380\times$ greater than the estimated scatter. Since the average 
linewidth of the detectors is around $Q_r \sim 20,000$, we can conclude that the 
resonant frequencies will not collide after irradiation to a flightlike dose.

Since the observed shift in the mean internal quality factor was positive, we conclude 
that the irradiation had no meaningful effect on the $Q_i$ values of the KIDs. It 
would be expected that more disorder in either the Al inductor or the Nb capacitor 
would cause more losses, decreasing $Q_i$.

\section{Conclusion}

In this work, we developed a cryogenic irradiation system to provide displacement 
damage doses representative of the energetic particle environment in Earth-Sun 
L2 orbit. We used this setup to irradiate an array of KIDs developed for the 
FIRESS instrument of PRIMA to approximately $62\%$ of the predicted 
mission dose, finding no significant changes in either the
quasiparticle lifetimes ($\tau_{qp}$), resonant frequencies ($f_r$), or internal quality factors ($Q_i$). 
Future work will involve irradiating the array to $100\%$ or more of 
the full mission dose, and taking measurements 
of $\tau_{qp}$, $f_r$, and $Q_i$ at multiple irradiation levels along the way.
This will allow us to conclusively determine the effects of displacement damage 
on $\tau_{qp}$, $f_r$ and $Q_i$.

\vfill

\end{document}